\documentclass[doublecol,preprint]{epl2nodraft}
\usepackage{textcomp}
\usepackage{amsmath}
\usepackage{amssymb}
\usepackage{graphicx}

\usepackage{color}
\usepackage{ulem}
\usepackage{ textcomp }

\begin{document}

\title{Elastic precession of electronic spin states in interacting integer quantum Hall edge channels} 

\author{Davide Venturelli$^{1,2}$ and Denis Feinberg$^{1}$}

\institute{$^{1}$Institut NEEL, CNRS and Universit\'{e} Joseph Fourier, Grenoble, France, \\
$^{2}$NEST, Scuola Normale Superiore and Istituto Nanoscienze-CNR,  Piazza San Silvestro 12, I-56127 Pisa, Italy\\}


\abstract{
We consider the effect of Coulomb interactions in the propagation of electrons, prepared in arbitrary spin states, on chiral edge channels in the integer quantum Hall regime. Electrons are injected and detected at the same energy at different locations of the Hall bar, which is modeled as a chiral Tomonaga-Luttinger liquid. The current is computed perturbatively in the tunneling amplitudes, within a non-crossing approximation using exact solutions of the interacting Green's functions. In the case of different channel velocities, the spin precession effect is evaluated, and the role of interaction parameters and wavevectors is discussed.
}
\maketitle


\section{Introduction} During the last few years the properties of chiral and adiabatic transport of Integer Quantum Hall edge states (IQHES) in 2-Dimensional Electron Gases have been employed for implementation of quantum interferometers: devices meant to test quantum coherence and to be possibly exploited for quantum information purposes~\cite{Exp1,Exp3}.

Recent results on the control over transport through quantum dots (QD) allowed exceptional experimental opportunities such as time-resolved single-electron injection~\cite{OnDemand}, or energy-resolved emission~\cite{EnergyEmission} and detection~\cite{Altimiras} on IQHES. In addition to the control of single charge transport, static spin manipulation on spin-resolved IQHES has been experimentally shown using arrays of micromagnets~\cite{Karmakar}, and there is a theoretical proposal to rotate spin states using spin-orbit coupling~\cite{DattaDasSOI}. This preparation of electronic wavepackets of arbitrary spin state (i.e. delocalized over spin-resolved channels) in IQHES clears the path towards the realistic implementation of the first controlled mesoscopic spin interferometer. Spin-interferometry is a famous example of fundamental coherence effects in quantum mechanics, which so far has been experimentally controlled only in neutron time-of-flight experiments, despite some theoretical proposals in mesoscopic systems~\cite{SpinInterf3}.  

In this work we theoretically analyze energy-resolved and spin-resolved transport between two tunnel contacts over a finite region of an interacting spinful chiral electron liquid, effectively studying a problem of spin interference. While the essential physics of IQHES is captured by a single-particle scattering approach, considering Coulomb interactions is essential in order to describe the energy-dependence of transport observables~\cite{Exp4anomalous} and dephasing at low temperature~\cite{AltimirasFrozen}.

We study interaction effects in the framework of the spinful chiral Tomonaga-Luttinger model (CTLM), which exhibits remarkable non-Fermi-liquid behavior such as fractionalization of excitations. This model for IQHES is justified on general theoretical basis~\cite{Zulike}, it is directly corroborated by several experimental hints, and it has been recently successfully employed to provide elements of understanding of the energy dependence of visibility in electronic Mach-Zehnder interferometers~\cite{Sukhorukov} and for non-equilibrium edge state equilibration~\cite{PDG1}.

\section{The transport model} Given the previous considerations, we can write the Hamiltonian of an interacting Hall bar in second quantization as a sum of a free part $H^{0}_{edge}$ and an interacting part $H^{int}_{edge}$. The free Hamiltonian features two quasi-particle modes with linear dispersions and non-zero difference between their Fermi velocities $v_{F}^{\sigma}$=$v_{F}+\sigma\delta v$ and their wavevectors $k_{F}^{\sigma}$=$k_{F}+\sigma\delta k$
\begin{equation}
H^{0}_{edge}=\sum_{\sigma,k}\hbar v_{F}^{\sigma}(k-k_{F}^{\sigma})a_{\sigma}(k)^{\dagger}a_{\sigma}(k)\ ,\label{eq:H0Luttinger}
\end{equation}
where $\sigma=\uparrow,\downarrow\equiv(+1,-1)$ and the $a_{\sigma}(k)$/$a_{\sigma}^{\dagger}(k)$ are operators which annihilate/create electrons with a given momentum $k$ and spin $\sigma$. This spin dependence of the parameters has a crucial effect that can be traced back to the entanglement between the orbital and spin degrees of freedom in IQHES: the two spin components acquire a phase difference during propagation due to the channel asymmetry. By measuring the spin state after propagation, this effect results in a spin precession around the axis of quantization (see fig.1c), whose period in space $\lambda_{X}$ depends on the energy $\epsilon$ of the particle:
\begin{equation}
\lambda_{X}=\pi\left[\left(\delta k+\epsilon\delta v/\left(v_{F}^{2}-\delta v^{2}\right)\right)\right]^{-1}\ .\label{eq:precessionperiod}
\end{equation}

Coulomb interactions between the two chiral 1D channels are included in the framework of the CTLM as local density-density terms between electrons of the same spin ($w$) or with different spin ($u$) ($g_{4\parallel}$ and $g_{g\perp}$ terms in the usual g-ology notation~\cite{Solyom})
\begin{equation}
H^{int}_{edge} \propto \sum_{\sigma,k>0}\left[w\;\rho_{\sigma}(k)\rho_{\sigma}(-k)+u\;\rho_{\sigma}(k)\rho_{-\sigma}(-k)\right] ,\label{eq:HINTLuttinger}
\end{equation}
where $\rho_{\sigma}(k)=\sum_{q}a^{\dagger}_{\sigma}(q)a_{\sigma}(q+k)$ are density fluctuation operators, which follow a Bose statistics. This form of the interaction allows to diagonalize the Hamiltonian by bozonisation, so to have analytical results for the propagators~\cite{LeeYangByczuk}. This is equivalent to a diagrammatics that can be resummed exactly~\cite{Dzyal} as we will exploit in the following. 

\begin{figure}\begin{centering}\includegraphics[width=\columnwidth]{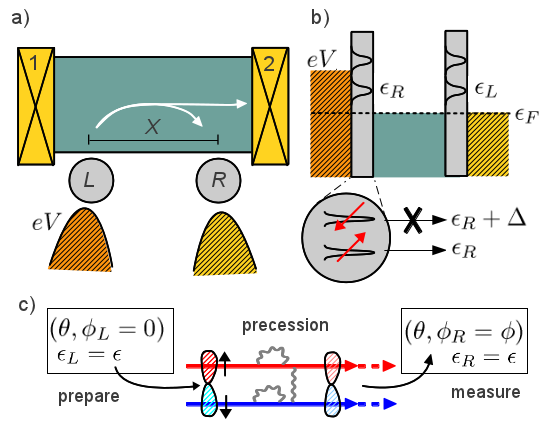}
\caption{(color online) a) Planar view of the four-terminal device, reservoirs are Ohmic contacts 1,2 and the shaded regions under the QDs L,R are the reservoirs. Contacts 1,2 are intended to be very far from L,R, ideally at infinity, and they are grounded. Only the reservoir connected to L is biased with respect to the ground. White arrows indicate the only non-compensated electronic currents. b) Energy level view of the transport experiment. Zoom inset: the tilted arrows represent the magnetic fields in the dots. c) Schematics of the transport experiment. Wiggled lines represent intra-edge interactions $w$ and inter-edge interactions $u$.\label{setupFIG}}\end{centering}\end{figure} 

The idealized setup (shown in fig.\ref{setupFIG}a) consists of two noninteracting QDs which are weakly tunnel-coupled to two spin-resolved IQHES. These are maintained at the same chemical potential $\epsilon_{F}$ by strong coupling with distant grounded ohmic contacts. QDs are in contact with non-interacting electron reservoirs at chemical potential $\mu_{R}$ and $\mu_{L}$ so that we can define a steady state tunneling current $I_{LR}$ from the left lead to the right lead, passing through the QDs $L$ and $R$, and the edge channel. The QDs are tuned so as to provide single resonant energy levels $\epsilon_{L}$ and $\epsilon_{R}$ in the transport window of the system (see fig.\ref{setupFIG}b), so that they can be described by the Hamiltonian $H_{i}^{d}=\epsilon_{i}d_{i}^{\dagger}d_{i}$, where $d_{i}$ are fermionic operators representing the charge occupation of the level of the QD $i={L,R}$. By setting $\mu_{R}=\epsilon_{F}$ and $eV>\epsilon_{F}$, because of chirality the only nonzero current in the system will be originated from dot L and will be drained either by the reservoir of dot R or by the distant ohmic contact 2. We are interested into the current passing through the right QD, defined as $I_{R}=e\frac{d}{dt}\left(d_{R}^{\dagger}d_{R}\right)=I_{LR}$.
By including the spin into the picture, the energy of the resonant levels $\epsilon_{i=R,L}$ becomes spin-resolved ($\epsilon_{i\uparrow}=\epsilon_{i\downarrow}+\Delta$) due to Zeeman energy and Coulomb charging energy so that a single spin projection is allowed in the transport window (say $\downarrow$), while transport of $\uparrow$ (possible in principle through cotunneling) is exponentially suppressed at low temperature and for large $\Delta$.
We now imagine that it is experimentally possible to orient the magnetic field acting on the dot along an arbitrary direction in space (defined by Euler's angles $\theta$ and $\phi$). The energy levels would be eigenstates with respect to the new spin direction $\left|\nearrow\right\rangle =\cos\left(\frac{\theta}{2}\right)\left|\uparrow\right\rangle +e^{i\phi}\sin\left(\frac{\theta}{2}\right)\left|\downarrow\right\rangle $; only an electron with spin state $(\theta_{i},\mbox{\ensuremath{\phi}}_{i})$ and energy $\epsilon_{i}$ would be allowed to tunnel through the QD and contribute to $I_{R}$. \textit{By fixing $\epsilon_{L}=\epsilon_{R}=\epsilon>\epsilon_{F}$ and $\theta_{L}=\theta_{R}=\theta$ this gedanken situation would define the spin-resolved, energy-resolved steady state transport problem object of this work.}

The complete setup is formally reproduced by the following model Hamiltonian
\begin{eqnarray}
H & = & H_{R}^{d}+H_{L}^{d}+H_{R}^{Res}+H_{L}^{Res}+H_{TR}+H_{TL}+\nonumber \\
 & + & H_{R}^{TE}+H_{L}^{TE}+H_{edge}\ ,\label{eq:total hamiltonian}
\end{eqnarray}
where the tunneling term transmitting a spin-projection through the aforementioned mechanism is written as
\begin{equation}
H_{i}^{TE}=t_{i}\left[\sum_{\sigma}V_{i}^{\sigma}\psi_{\sigma}(x_{i})\right]d_{i}^{\dagger}+h.c.\ ,\label{eq:Htunneledge}
\end{equation}
where $V_{i}^{\uparrow}=\cos\frac{\theta}{2}$ and $V_{i}^{\downarrow}=\sin\frac{\theta}{2}e^{i\phi_{i}}$
represent the spin projection in the direction fixed by the fields in the QD, and $\psi_{\sigma}(x_{i})$ is the wavefunction of an electron at position $x_{i}$ in the IQHES, Fourier transform of $a_{\sigma}(k)$ operators defined in $H_{edge}$ previously discussed. The reservoir Hamiltonians $H_{i}^{Res} = \sum_{k}\left(\epsilon_{k}-\mu_{i}\right)c_{k}^{\dagger}c_{k}$ and $H_{Ti} = \sum_{k}\left(T_{k}c_{k}^{\dagger}d_{i}+T_{k}^{\star}d_{i}^{\dagger}c_{k}\right)$ are just free Fermi liquids (electronic operators $c_{k}$), tunnel coupled (by means of tunnel amplitudes labeled $T_{k}$) to the QDs.

We need to point out that within the present technology, it is not realistic to have local and strong magnetic fields such as to define a proper arbitrary basis in the QDs. In our setup such strong static fields have been introduced aiming at a simple treatment of the electronic transport. Injection and detection of single electrons over co-propagating IQHES is also a subtle technical point, as standard quantum point contacts usually work in the adiabatic transmission regime. We expect nevertheless that the desired coupling might be achievable by employing suspended tunneling tips over surface electron gases~\cite{morgenstern} or perhaps with single-electron turnstiles~\cite{Battista}. We also note that it is now possible to prepare locally an electronic spin state in QDs~\cite{Meunier1}, and to measure its spin state by means of spin-dependent tunneling rates~\cite{Spinreadout1,Meunier2}. Recently it has also been shown in experiments that single-electron manipulation in a QD can in principle be followed by propagation of the single electron over a chiral 1-dimensional channel and by detection and trapping in a second QD~\cite{QubitTransfer12}. Concerning the charge degree of freedom, our setup could be thought as the steady-state version of these experiments, using IQHES instead of surface acoustic waves in order to enforce the chirality of transport.
Keeping in mind these technological considerations, we can assess our transport calculation as an approximation of a time-dependent transport procedure hopefully achievable in the near future.%

\section{Calculation of the current} We will now compute the current $I_{R}$ in lowest-order perturbation theory in the tunneling amplitudes $t_{L}$ and $t_{R}$. Following common non-equilibrium Green's functions (GF) approaches for tunneling in mesoscopic devices we write the steady-state current through the detector QD (at position $x_{R}$) as
\begin{equation}
I_{R}=-\frac{e}{h}\left|t_{R}\right|^{2}\sum_{\sigma\sigma^{\prime}}V_{\sigma}^{R}V_{\sigma^{\prime}}^{R*}\int\frac{d\omega}{2\pi}[F_{\sigma\sigma^{\prime}}^{<}(x_{R},x_{R};\;\omega)g_{R}^{>}(\omega)]\ .\label{eq:currentformulajahuo}
\end{equation}
where $F_{\sigma\sigma^{\prime}}^{<}(x_{R},x_{R};\;\omega)$ is the time Fourier transform of the GF $\langle\psi_{\sigma}^{\dagger}(x_{R},t)\psi_{\sigma^{\prime}}(x_{R},t^{\prime})\rangle$, and $g^{>}_{R}(\omega)$ is the analogous GF of the QD operators. By means of the fluctuation-dissipation theorem,  $g^{>}_{R}(\omega)$ is proportional to the product of the inverse QD occupation function $1-f_{R}(\omega)$ and a sharp density of states $\rho_{R}\delta(\omega-\epsilon\hbar^{-1})$ (note that $g^{<}_{R}(\omega)=0$ for $\omega>\epsilon_{F}$ in our setup, since $f_{L}(\omega>\epsilon_{F})=0$).
$F_{\sigma\sigma^{\prime}}^{<}$ must be expanded at order $|t_{L}|^{2}$ in order to take into account the effects of the injection at position $x_{L}$. We choose to compute this expansion by observing that $F_{\sigma\sigma^{\prime}}^{<}(x_{R},x_{R};\;\omega)$ can be in principle obtained from the time-ordered GF on the Keldysh contour $G^{T}=\langle\mathcal{T}\psi^{\dagger}(x_{R},\tau)\psi(x_{R}^{\prime},\tau^{\prime})\rangle$ by analytical continuation (Langreth's Theorem)~\cite{CAROLINOZIERES}.
In terms of the total self-energy of the system $\Sigma_{tot}^{<}(\bar{x},\bar{x}^{\prime};\omega)$, the steady-state quantum kinetic equations take the standard form~\cite{HaugJauho}
\begin{equation}
G^{<}(x_{R},x_{R}) = \int{d\bar{x}d\bar{x}^{\prime}}G^{+}(x_{R},\bar{x})\Sigma_{tot}^{<}(\bar{x},\bar{x}^{\prime})G^{-}(\bar{x}^{\prime},x_{R})\ ,\label{eq:Langreth rules}
\end{equation}
where $G^{+}/G^{-}$ denote \textit{fully-dressed} retarded/advanced GFs (we omitted the $\sigma$ ad $\omega$-dependence of the functions). These functions can be easily evaluated since they obey an equilibrium Dyson equation. They do not get renormalized by tunneling, as chirality implies that $G^{-}$($G^{+}$) is zero if evaluated for positive(negative) distances, so they are the fully-interacting GFs of the CTLM, which we denote $\mathcal{G}^{+/-}$.
\begin{figure}
\begin{centering}
\includegraphics[width=\columnwidth]{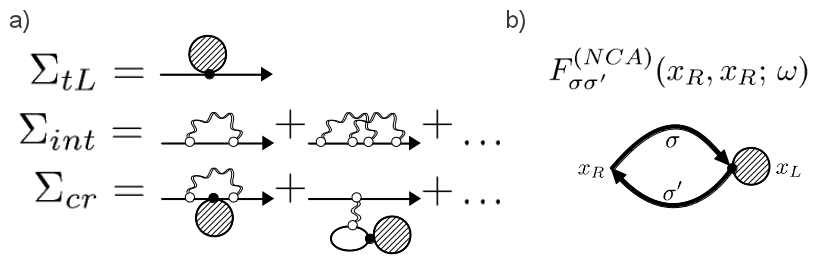}
\caption{a) $\Sigma_{tL}$ is a point-like elastic diagram in x-$\omega$ representation. $\Sigma_{int}$ can be expanded in diagrams containing the interaction propagators (both $u$ and $w$ interactions contribute, represented as wiggle lines). $\Sigma_{cr}$ contain both kind of vertices. b) Expansion of $F_{\sigma\sigma^{\prime}}^{<}(x_{R},x_{R};\;\omega)$ at order $|t_{L}|^{2}$ with the NCA. Thick fermionic lines represent $\mathcal{G}$, dressed by $\Sigma_{int}$ 
\label{NCAfig}}
\end{centering}\end{figure}
All the difficulty lies in evaluating the total self-energy which, in terms of diagrams, is derived through the irreducible (with respect to a cut of fermionic lines)  contributions obtainable with \textit{both} the interaction vertices $w$ and $u$ and $t_{L}$.
This is a cumbersome task, as the exact result consists of difficult time-integrations and Fourier-tranforms of combinations of two-particle correlation functions in time and space domain, as already recognized in similar two-site tunneling problems~\cite{AVERINNAZAROVSOTAKEI}. We decide then to employ a non-crossing approximation (NCA) on the total self-energy, by stating that $\Sigma_{tot}\simeq\left(\Sigma_{tL}+\Sigma_{int}\right)$.
The first part of the self-energy is the tunneling self-energies due to the coupling to the QD at $x_{L}$. In particular, $\Sigma_{tL}^{<}(\sigma,\sigma^{\prime},\omega)$=$V_{\sigma}^{L}V_{\sigma^{\prime}}^{L*}g_{L}^{<}(\omega)$~\cite{HaugJauho}. The second term is the self-energy due to Coulomb interactions. The exact solution with $\Sigma_{int}$ alone os given by the bosonization and the GF $\mathcal{G}^{+/-}$. The NCA consists in neglecting a third term $\Sigma_{cr}$ which represents the crossed diagrams between the tunneling and the interaction vertices (see fig.\ref{NCAfig}a).
If this approximation is employed, the final result takes the form of the noninteracting Fisher-Lee formula for the current where the retarded GF are now fully interacting
\begin{equation}
I_{R}=\frac{e}{h}\left|t_{R}\right|\left|t_{L}\right|^{2}\rho_{R}\rho_{L}\sum_{\sigma^{\prime}\sigma}\Gamma_{\sigma^{\prime}\sigma}\mathcal{G}_{\sigma^{\prime}}^{+}(X;\epsilon)\left[\mathcal{G}_{\sigma}^{+}(X,\epsilon)\right]^{*}\label{eq:finalcurrentelastic}
\end{equation}
where $X=x_{L}-x_{R}$ (see fig.\ref{NCAfig}b) and $\Gamma_{\sigma^{\prime}\sigma}=V^{\sigma}_{R}(V^{\sigma^{\prime}}_{R})V^{\sigma^{\prime}}_{L}(V^{\sigma}_{L})^{*}$. We note that in the inelastic regime ($\epsilon_{L}$ $\neq$ $\epsilon_{R}$) the NCA would not give any contribution: the energy-relaxing terms are not captured in this approximation.
We also note that the recovery of a Landauer-B\"uttiker-like formula for the current is an exceptional case which is usually never valid for many-body treatment of Coulomb interacting systems, as it has been shown extensively for transport through QDs~\cite{NESS}. Conceptually the NCA consists in neglecting the (virtual) interplay of injection/detection and interactions, and to consider its effect only on the propagation region, in which is similar to other approximations recently used for interacting electrons in interferometers or quantum wires~\cite{CHALKERKOVPDG2}. Although in the CTLM for local interactions the evaluation of some corrections is possible (we can show that all correction diagrams are small for sufficiently long distances), ultimately the justification of eq.\ref{eq:finalcurrentelastic} for extended propagation regions should lie in the chirality of the system, in the filtering of the elastic signal, in the weak-tunneling regime, but it is independent on the model of the interactions, as long as they are sufficiently local.

In order to compute the currents, we turn to the CTLM with local interactions. We analytically found by direct integration of the well-known GF in space/time~\cite{Simon} the Fourier transforms of $\mathcal{G}_{\sigma}^{<}\left(x,\omega\right)$ for positive energies. In terms of Confluent Hypergeometric Functions (sending all regularization cut-offs to zero) we obtained
\begin{equation}
\mathcal{G}_{\sigma}^{>}\left(x,\omega\right)=\frac{i\Theta\left(\omega\right)}{2\pi}e^{i[k_{F}^{\sigma}+\omega(\frac{1}{v_{-}})]x}\Phi[\frac{1}{2}+\sigma\delta,1,ix\omega(\frac{1}{v_{+}}-\frac{1}{v_{-}})]\label{eq:AnalyticalConfluentG}
\end{equation}
where $\Phi\left(a,b,z\right)$ is Kummer's function~\cite{STEGUN} and the velocities of the new bosonic eigenmodes are $v_{\pm}=\bar{v}+\frac{w}{\pi}\pm\frac{1}{2}\sqrt{4\delta v^{2}+\frac{u^{2}}{\pi^{2}}}$, where $\bar{v}$=$(v_{F}^{\uparrow}$+$v_{F}^{\downarrow})/2$ and $\delta$=$\sin[\arctan(\pi\delta v/u)]/2$ encodes the spin-asymmetry. For $\delta=0$ the GF reduces to a Bessel function featuring exact spin-charge separation and for $\delta=1/2$ ($u$=0) we obtain two spin-decoupled GF, i.e. independent spin-polarized Fermi liquids.
The spin dependence enters as a phase of $\mathcal{G}^{>}(x,\omega)$. In the analytical form of $\Phi$ there is a functional equivalence between $\omega$ and $x$, so its algebraic decay is the same for energy and distance. 

\begin{figure}
\begin{centering}
\includegraphics[width=\columnwidth]{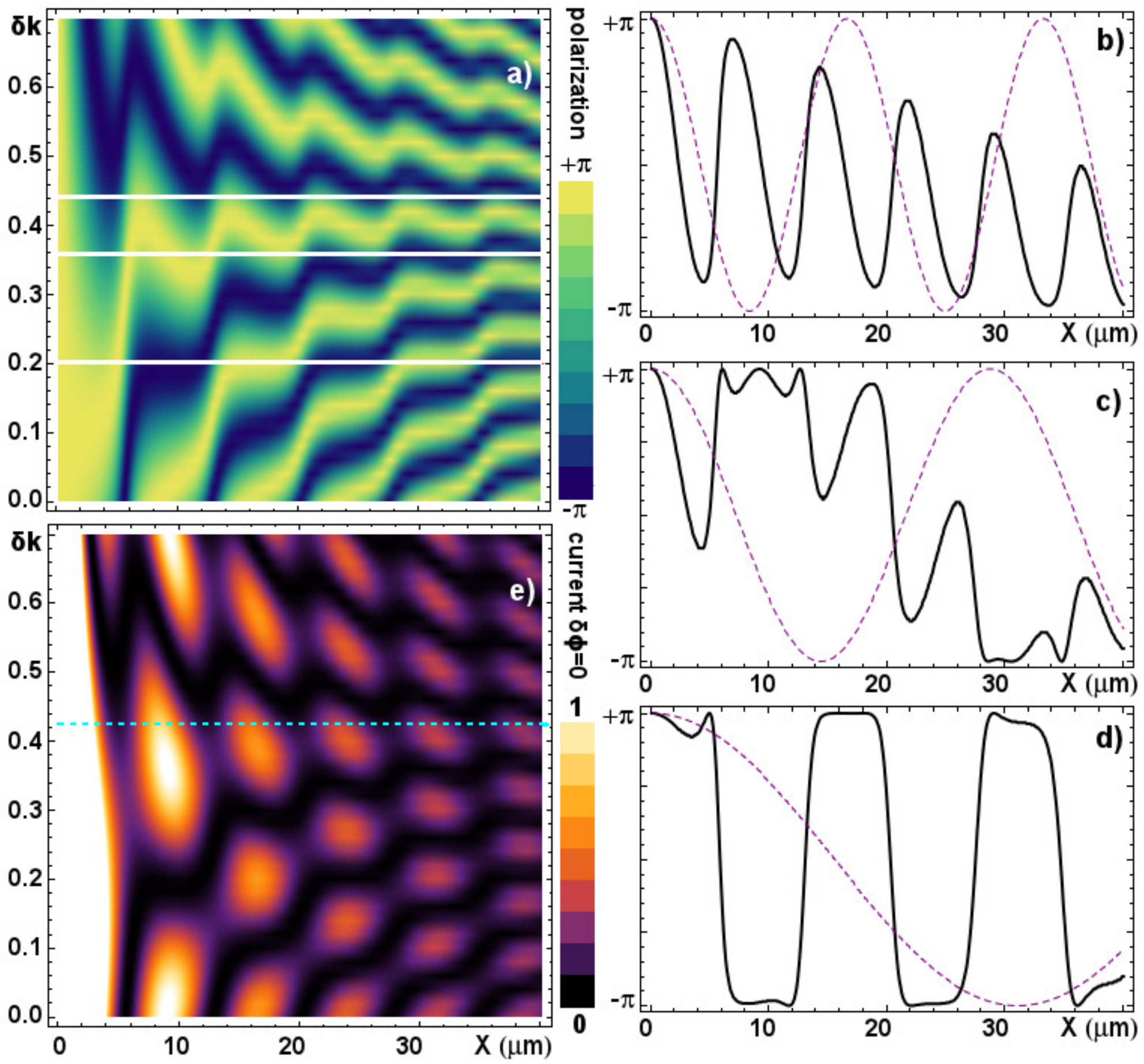}
\caption{(color Online) a) maximum phase precession angle $\delta \phi$ as a function of $\delta k$ and X. b-d) represents cuts of plot (a) for different wavevectors (thick line). The dashed line indicate the non-interacting precession evolution whose period is described by eq.\ref{eq:precessionperiod}. e) current signal for $\delta\phi=0$ in the measurement setup, corresponding to filtering the ``up'' spin projection in the $\theta=\pi/2$ basis. The dashed line indicates the condition $\lambda_{X}=\infty$ using eq.\ref{eq:precessionperiod} with the eigenmode velocities. \label{GFfig}}
\end{centering}\end{figure}
\section{Spin precession} The velocities of spin-resolved IQHES have been measured to be of the order of $10^{5}ms^{-1}$, and the propagation distances for coherent transport in nanostructures can be as high as hundreds of microns~\cite{Exp3}.
It is very hard to have a realistic idea on what might be $\delta v$ in interacting models, as the bands are heavily influenced by the interactions~\cite{DEMPSEYZOZOUinash}. We note however that the use of top gates for selective addressing and manipulation of IQHES necessarily separates in space the edge trajectories of the two spin-components. In the adiabatic non-interacting limit $\delta v$ can be estimated to be roughly proportional to second derivative of the confinement potential, i.e. non-zero for effective smooth confinements which are at least parabolic. In the following, we will consider the test-case of a deviation of about $5\%$ from the average value $\bar{v}$.
Estimating the CTLM parameters of intra-edge interaction $w$ and inter-edge interaction $u$ is another important problem in interacting IQHES, object of intense theoretical and experimental research. They are likely to be of the same order of magnitude~\cite{Sukhorukov} but their value depends on details of the nanostructures such as the screening by the metallic surrounding. As reasonable working conditions we take $w\simeq u\simeq 0.7v_{F}$, zero temperature (i.e. $k_{B}T$ $\lesssim$ $10^{-6}eV$, corresponding to a few tens of mK) and $\epsilon_{max}$ $\lesssim$ $\hbar\omega_{C}\simeq10^{-3}eV$ in order to justify linear dispersion.

As an example we fix $\theta=\pi/2$, which means that we are symmetrically coupled to both spin channels on the Hall bar (i.e.,
we are transporting spin qubits on the cartesian x direction on the Bloch sphere). The transmission coefficient of this spin state will oscillate in $\delta\phi$ as
\begin{equation}
T_{X}\left(\epsilon\right)=\frac{1}{2}[M_{X}\left(\epsilon\right)+\cos(\delta\phi)\mbox{Re}F_{X}\left(\epsilon\right)-\sin(\delta\phi)\mbox{Im}F_{X}\left(\epsilon\right)]\label{eq:Tx}\end{equation}
where $M_{X}\left(\omega\right)$ = $\left|\mathcal{G}_{\sigma}^{R}\left(X,\omega\right)\right|^{2}$
and $F_{X}\left(\omega\right)$ = $\mathcal{G}_{\downarrow}^{R}\left(X,\omega\right)\mathcal{G}_{\uparrow}^{R}\left(X,\omega\right)^{*}$.


Similarly to the tuning of the arm lengths in a Mach-Zehnder interferometer, by filtering the proper precession angle (which is a function of $\epsilon$ or $X$) we could maximize the current signal up to reaching the transmission of a single spin-component along the z-axis (max $T_X(\epsilon)=M_X(\epsilon)$).
Although the phase acquired by means of wavevector difference has just an additive interplay with the energy-dependent phase-shift their independent evolution generates a non-trivial precession in space. For instance, by replacing $v_{F}^\sigma$ with the new velocities  $v_{\pm}$ in eq.\ref{eq:precessionperiod} we obtain indeed a new condition for which the precession could be frozen in principle ($\delta k=\omega (v_{+}-v_{-})/(v_{+}v_{-})$). However, for $u\neq0$ the phase difference cannot be locked as in the non-interacting limit, since $M_{X}(\epsilon)$ necessarily oscillates in space independently from the spin-asymmetry. More generally we might end up with a smoothly regular, irregular or sharply-varying precession, which can be visualized by  plotting the $\phi$-angle for which the transmission probability is maximum as a function of distance, as in fig.~\ref{GFfig}a-d. Besides the spectroscopic utility of this peculiar pattern, in this slanted basis such rich behaviour could have very practical application in the context of elastic flying spin-qubit transport on edge channels. With proper tuning of $\delta k$~\cite{Karmakar} we can in fact obtain 'square-wave'-like spin-flip transitions (see fig.~\ref{GFfig}b), implying that the spin-projection of the electron can be deterministically defined over hundreds of nanometers.

\section{Conclusions} In conclusion, we studied a gedanken experiment with IQHES subject to Coulomb interactions, where difference in Fermi velocities enables a spin precession effect. We developed a non-crossing approximation in order to compute the current relative to the elastic propagation of a spin state as a function of energy and distance. Understanding the precession of spin states complements the general value of non-local tunneling spectroscopy for analyzing the Coulomb interaction features of transport in interacting IQHES and might result in useful application for flying spin-qubits architectures.

\acknowledgments
We acknowledge useful discussions with P. Degiovanni, V. Giovannetti, M. di Ventra. This work was by the Italian MIUR under the FIRB IDEAS project RBID08B3FM and financial support from EU through the project GEOMDISS.


\begin{thebibliography}{99}

\bibitem{Exp1}Yang Ji, Yunchul Chung, D. Sprinzak, M. Heiblum,
D. Mahalu, H. Shtrikman, Nature \textbf{422} (2003) 415; I. Neder, N. Ofek, Y. Chung,
M. Heiblum, D. Mahalu, V. Umansky, Nature \textbf{448} (2007) 333.

\bibitem{Exp3}P. Roulleau, F. Portier, P. Roche, A. Cavanna,
G. Faini, U. Gennser, and D. Mailly, Phys. Rev. Lett. \textbf{100} (2008) 126802.

\bibitem{OnDemand}G. Feve, A. Mahe, J.-M. Berroir, T. Kontos,
B. Pla\c{c}ais, D. C. Glattli, A. Cavanna, B. Etienne and Y. Jin, Science \textbf{316} (2007) 1169

\bibitem{EnergyEmission}C. Leicht, P. Mirovsky, B. Kaestner, F. Hohls, V. Kashcheyevs, E. V. Kurganova, U. Zeitler, T. Weimann, K. Pierz and H. W. Schumacher, Semicond. Sci. Technol. \textbf{26} (2011) 055010.

\bibitem{Altimiras}C. Altimiras, H. le Sueur, U. Gennser, A. Cavanna,
D. Mailly, and F. Pierre, Nature Physics \textbf{6},
(2009) 34.

\bibitem{Karmakar}B. Karmakar, D. Venturelli, L.
Chirolli, F. Taddei, V. Giovannetti, R.Fazio, S.
Roddaro, G. Biasiol, L. Sorba, V. Pellegrini, and F.
Beltram, Phys. Rev. Lett. \textbf{107} (2011) 236804.

\bibitem{DattaDasSOI}L. Chirolli, D. Venturelli, F. Taddei, Rosario Fazio, V. Giovannetti arXiv:1111.0675v2 preprint 2012.

\bibitem{SpinInterf3}P. Simon and D. Feinberg, Phys. Rev. Lett. \textbf{97} (2006) 247207; Radu Ionicioiu and Irene D\textquoteright{}Amico,
Phys. Rev. B \textbf{67} (2003) 041307(R).

\bibitem{Exp4anomalous}I. Neder, M. Heiblum, Y. Levinson, D. Mahalu, and V. Umansky, Phys. Rev. Lett. \textbf{96} (2006) 016804.

\bibitem{AltimirasFrozen}C. Altimiras, H. le Sueur, U. Gennser, A. Cavanna, D. Mailly, and F. Pierre, Phys. Rev. Lett. \textbf{105} (2010) 226804.

\bibitem{Zulike}U. Z\"ulicke, A. H. MacDonald, Phys. Rev. B \textbf{54} (1996) 16813.

\bibitem{Sukhorukov}I. P. Levkivskyi and E. V. Sukhorukov, Phys. Rev. B \textbf{78} (2008) 045322.

\bibitem{PDG1}P. Degiovanni, Ch. Grenier, G. Feve, C. Altimiras,
H. le Sueur, and F. Pierre, Phys. Rev. B \textbf{81} (2010) 121302(R)

\bibitem{Solyom}J. Solyom, Advances in Physics, vol.\textbf{28} (1979) 201.

\bibitem{LeeYangByczuk} H. C. Lee and S.-R. Eric Yang, Phys. Rev. B \textbf{56}, R15529 (1997); K. Byczuk, Phys. Rev. B \textbf{57} (1998) 3821.

\bibitem{Dzyal}I. E. Dzyaloshinskii and A. I. Larkin, Zh. Eksp. Teor. Fiz. \textbf{65} (1973) 411.

\bibitem{morgenstern} S. Becker, C. Karrasch, T. Mashoff, M. Pratzer,
M. Liebmann, V. Meden, and M. Morgenstern, Phys. Rev. Lett. \textbf{106} (2011) 156805.

\bibitem{Battista} F. Battista, P. Samuelsson, arXiv:1112.4286 preprint, 2012.

\bibitem{Meunier1}F. H. L. Koppens, C. Buizert, K. J. Tielrooij,
I. T. Vink, K. C. Nowack, T. Meunier, L. P. Kouwenhoven and L. M.
K. Vandersypen, Nature \textbf{442} (2006) 766.

\bibitem{Spinreadout1}J. M. Elzerman, R. Hanson, L. H. Willems van Beveren,
B. Witkamp, L. M. K. Vandersypen \& L. P. Kouwenhoven, Nature
\textbf{430} (2004) 431.

\bibitem{Meunier2}R. Hanson, L. H. Willems van Beveren, I.
T. Vink, J. M. Elzerman, W. J. M. Naber, F. H. L. Koppens, L. P. Kouwenhoven,
and L. M. K. Vandersypen, Phys. Rev. Lett. \textbf{94} (2005) 196802. 

\bibitem{QubitTransfer12}S. Hermelin, S. Takada, M. Yamamoto, S. Tarucha, A. D. Wieck, L.Saminadayar, C. B\"auerle, Tristan Meunier, Nature \textbf{477} (2011) 435; R. P. G. McNeil, M. Kataoka, C. J. B. Ford, C. H. W. Barnes, D. Anderson, G. A. C. Jones, I. Farrer, D. A. Ritchie, Nature \textbf{477} (2011) 439 

\bibitem{CAROLINOZIERES}C. Caroli, R. Combescot, P. Nozieres and D. Saint-James, J. Phys. C \textbf{5} (1971) 21

\bibitem{HaugJauho}H.J.W. Haug, A. Jauho, Quantum Kinetics in Transport
and Optics of Semiconductors, 2nd Edition, Springer (2008)

\bibitem{AVERINNAZAROVSOTAKEI} D. V. Averin and Yu. V. Nazarov, Phys. Rev. Lett. \textbf{65}, 2446 (1990); S. Takei, M. Milletarì, and B. Rosenow, Phys. Rev. B \textbf{82}, (2010) 041306(R). 

\bibitem{NESS}H. Ness, L. K. Dash, and R. W. Godby, Phys. Rev. B \textbf{82} (2010) 085426.

\bibitem{CHALKERKOVPDG2}D. L. Kovrizhin and J. T. Chalker, Phys. Rev. B \textbf{80} (2009) 161306; Ch. Grenier, R. Herv\'e, G. F\`eve,
P. Degiovanni, Mod. Phys. Lett. B \textbf{25} (2011) 1053; H. Aita, L. Arrachea and C. Naon, J. Phys. Condens. Matter \textbf{23} (2011) 475601.

\bibitem{Simon}B. Braunecker, C. Bena, and P. Simon, Phys. Rev. B \textbf{85} (2012) 035136. 

\bibitem{STEGUN}M. Abramowitz, I. A. Stegun - {}``Handbook of Mathematical
Functions'' - New York: Dover Publications (1972)

\bibitem{DEMPSEYZOZOUinash}J. Dempsey, B. Y. Gelfand, and B. I. Halperin, Phys. Rev. Lett. \textbf{70} (1993) 3639; S. Ihnatsenka and I. V. Zozoulenko, Phys. Rev. B \textbf{73} (2006) 155314.


\end{thebibliography}
\end{document}